\documentclass[12pt ]{iopart}
\usepackage{amsbsy,amssymb}
\usepackage{epsfig}

\date{} 

\begin{document}

\title[Enhanced Dynamo Drive for RFP Sawtooth Relaxation Process]{Enhanced dynamo drive for the sawtooth relaxation process due to non-uniform resistivity distribution  in a  reversed field pinch}

\author{Wentan Yan$^{1}$, Ping Zhu$^{2,3}$, Hong Li$^{1}$, Wandong Liu$^{1}$, Bing Luo$^{4}$, Haolong Li$^{5}$}
\address{$^1$ KTX Laboratory and Department of Plasma Physics and Fusion Engineering, University of Science and Technology of China, Hefei, Anhui 230026, China}
\address{$^2$  State Key Laboratory of Advanced Electromagnetic Engineering and Technology, International Joint Research Laboratory of Magnetic Confinement Fusion and Plasma Physics, School of Electrical and Electronic Engineering, Huazhong University of Science and Technology, Wuhan, Hubei 430074, China}
\address{$^3$ Department of Nuclear Engineering and Engineering Physics,University of Wisconsin-Madison, Madison, Wisconsin 53706, USA}
\address{$^4$ Hunan Province Key Laboratory Integration and Optical Manufacturing Technology, College of Mathematics and Physics, Hunan University of Arts and Science, Changde, Hunan 415000, China}
\address{$^5$  Department of Physics, School of Science, Tianjin University of Science and Technology, Tianjin 300457,  China}
\ead{zhup@hust.edu.cn, honglee@ustc.edu.cn}
\vspace{10pt}
\begin{indented}
\item[] Date: 20 February 2025
\end{indented}

\begin{abstract}
In this work, we use the three-dimensional resistive MHD code NIMROD to investigate the impact of resistivity inhomogeneity on the sawtooth process of  an  reversed field pinch (RFP) plasma. The simulation employs a non-uniform resistivity profile similar to experiments, which monotonically increases from the core to the edge as the temperature decreases. The resistivity inhomogeneity introduces an additional electric field  in the plasma, which  accelerates the inward diffusion of magnetic flux and changing the self sustained reversal state, hence significantly enhances the dynamo effect  and the sawtooth process in the  RFP plasma.
\end{abstract}

\maketitle

\section{Introduction}
RFP (Reversed Field Pinch) \cite{Marrelli-2021} is a toroidal  magnetic configuration for plasma confinement, where the toroidal  and the poloidal components of magnetic field are of the same order of magnitude. The strong shear of the poloidal field stabilizes the ideal kink instability in the RFP, leading to the dominance of tearing mode instabilities with $m=-1, n\sim 2R/a$. Due to the absence of a strong guiding magnetic field, the core resonant region where multiple tearing modes coexist is often in a state of magnetic stochasticity, resulting in a rich collection of self-organization behaviors within the RFP plasma \cite{Gupta-2023}, which often involve the dynamo effect. A dynamo electric field converts the externally applied poloidal magnetic flux to toroidal magnetic flux, maintaining the poloidal current and the toroidal field reversal near plasma boundary. The dynamo effect is also known as the flux pumping in the context of tokamaks, which has been used to explain the physical mechanism of the sawtooth-free steady state \cite{Jardin-2015,Jardin-2020,Krebs-2017,Piovesan-2017,Yu-2024}. In the rising and crash phases of a RFP sawtooth oscillation, the dynamo field exhibits different characteristics. During the  rising phase, the plasma current profile gradually peaks, and multiple $m=-1$ tearing modes  saturate and  interact nonlinearly among each other \cite{Tamano-1987}.  When the nonlinear interaction  level reaches a threshold, the plasma enters a rapid crash phase, the amplitude of the $m=-1$ tearing modes decay, and the nonlinear  beating leads to  modes such as the $(0,1)$ mode  growing rapidly. The dynamo electric field during the crash phase transport the magnetic flux from the core to the edge, and flatten the current profile to a relaxed state.

Nonlinear MHD simulations have long been conducted to explain the sawtooth phenomena in experiments and explore the underlying  physical processes \cite{Cappello-2000}.  Futch et al \cite{Futch-2018} and  Schnack  et al  \cite{Schnack-1985} discovered that when the pinch parameter $\Theta$ is larger, the field reversal surface and the resonance region of the tearing modes are closer, making the three-wave nonlinear interaction more likely to occur, and lead to the onset of sawtooth events. Kusano \cite{Kusano-1990} found in his 1990 simulation that in cases with a larger Lundquist number, the nonlinear effects of the plasma are enhanced and the sawtooth phenomena are more intense. In simulations by Ho and Craddock \cite{Ho-1991}, the quasilinear and  nonlinear dynamos  during the quasi-periodic sawtooth process were classified using the power flux diagnostic in Fourier space. King et al \cite{King-2011} and Sauppe and Sovinec \cite{Sauppe-2016,Sauppe-2017} studied the two-fluid effects  using NIMROD simulations, and the roles of Hall dynamo and MHD dynamo before and after the sawtooth burst were identified. Veranda et al \cite{Veranda-2020} used the resistivity with the Lundquist number $(S=10^5)$ closer to the experimental regime $(S\approx10^6\sim{}10^7)$  in simulations of the sawtooth process on RFX, and found the existence of current sheets in the sawtooth crash similar to those in experiments, confirming that the rapid magnetic reconnection process during the crash follows the modified Sweet-Parker scaling law.

Although ohmic heating and peaked temperature profiles have been proved to be significant factors affecting the  behavior of sawtooth in simulations and experiments \cite{Jardin-2015,Werley-1985,Vivenzi-2023}, which suggests the potential significance of resistivity, previous simulations often assume uniform resistivity for simplicity, and the role of resistivity and in particular its inhomogeneity in the sawtooth relaxation process in reality have been less clear. In this work we study the impact of resistivity inhomogeneity  on the sawtooth behavior in  three-dimensional MHD simulations using the NIMROD code \cite{Sovinec-2003} with various different resistivity profiles but  similar pinch parameter and Lundquist number. These simulations demonstrate how the non-uniform resistivity profile affects the three-dimensional mode evolution and the self-organization behavior of  plasma from the perspective of MHD dynamos.

The rest of this paper is organized as follows. Section 2 introduces  the MHD model in the NIMROD code and the simulation setups. Section 3 presents  the evolution of physical quantities in a typical  simulation of the sawtooth process. Section 4 examines the influence of non-uniform resistivity profiles on the  magnetic  diffusion. We also analyze how these profiles change the dynamo  sustaining the reversal in toroidal magnetic field. Finally a summary and discussion is  presented in Section 6.

\section{ Simulation model and setups}
We use the three-dimensional nonlinear  single-fluid MHD model implemented in the  NIMROD code  \cite{Sovinec-2003,Glasser-1999}  to simulate the sawtooth process in an RFP plasma.  The MHD equations are
\begin{eqnarray}
\frac{\partial N}{\partial t} &=& -\nabla\cdot(N\boldsymbol{v}) + \nabla\cdot(D_N\nabla N) \\
\rho\frac{\partial\boldsymbol{v}}{\partial t} &=& -\rho\boldsymbol{v}\cdot\nabla\boldsymbol{v} + \boldsymbol{J}\times\boldsymbol{B} - \nabla P + \rho\nu\nabla^2\boldsymbol{v} \\
\frac{\partial\boldsymbol{B}}{\partial t} &=& -\nabla\times E  \\
\frac{\partial T}{\partial t} &=& -\boldsymbol{v}\cdot \nabla T - T(\gamma-1)\nabla\cdot \boldsymbol{v} + \nabla\cdot(\chi \nabla T) \\
\boldsymbol{E} &=& -\boldsymbol{v}\times\boldsymbol{B} + \eta \boldsymbol{J}
\end{eqnarray}
while $N$, $\rho$, $\boldsymbol{v}$, $\boldsymbol{J}$, $\boldsymbol{B}$, $P$, $\boldsymbol{E}$, and $T$ represent plasma   number density,  mass density, velocity, current  density, magnetic field, pressure, electric field, and temperature, respectively. $D_n$, $\nu$ and $\chi$ are the coefficients for density  diffusivity, viscosity, resistivity, and isotropic thermal conductivity, respectively. $\eta$ is the resistivity and $\gamma$ is the adiabatic index. The dimensionless Lundquist number $S=\tau_R/\tau_A$ and Reynolds number $M=\tau_\nu/\tau_A$ characterize the resistive and viscous dissipations, where $\tau_A=a\sqrt{\mu_0 m_in}/B$ represents the Alfvén time scale, $\tau_R=\mu_0 a^2/\eta_0$ represents the resistive diffusion time scale, and $\tau_\nu=a^2/\nu$ represents the viscous dissipation time.

\begin{figure}[t]
\centering
\includegraphics[width=4in]{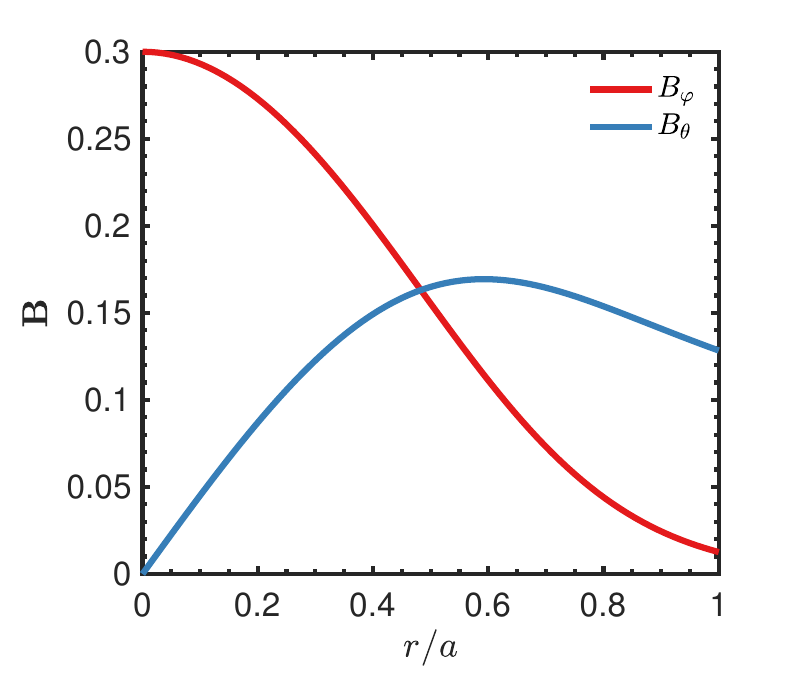}
\caption{The toroidal and the poloidal components of the magnetic field as  functions of the normalized  radius $r/a$ for the initial paramagnetic pinch equilibrium with $a\lambda(0) = 3.8$, where $a$ is the minor radius of plasma.}  \label{fig:ParaEq}
\end{figure}

A periodic cylinder  with  the KTX geometry is employed, where the major radius $R=1.4m$ and the minor radius $a=0.38m$ \cite{Liu-2019}.
 The simulations  start from a force-free equilibrium (Fig. \ref{fig:ParaEq}), which satisfies  the Ohmic steady-state condition, $\boldsymbol{E}_{eq} = -\boldsymbol{v}_{eq} \times \boldsymbol{B}_{eq} + \eta \boldsymbol{J}_{eq} = E_{loop} \boldsymbol{e}_\phi$, where a constant toroidal electric field $E_{loop}$ is applied at the boundary acting as the loop voltage, and the initial perturbation is  set with a relative amplitude of $\delta B/B = 10^{-5}$.  At the radial boundary,  the perturbed magnetic and velocity fields adhere to the ideal and solid-wall no-slip conditions, whereas the temperature perturbation complies with the Dirichlet condition. The initial equilibrium is one-dimensional and thus lacks a reversal at the boundary as per Cowling's theorem \cite{COWLING-1957}. The emergence of reversal in simulation  is the result of the three-dimensional self-organization and relaxation process.

\begin{figure}[t]
\centering
\includegraphics[width=4in]{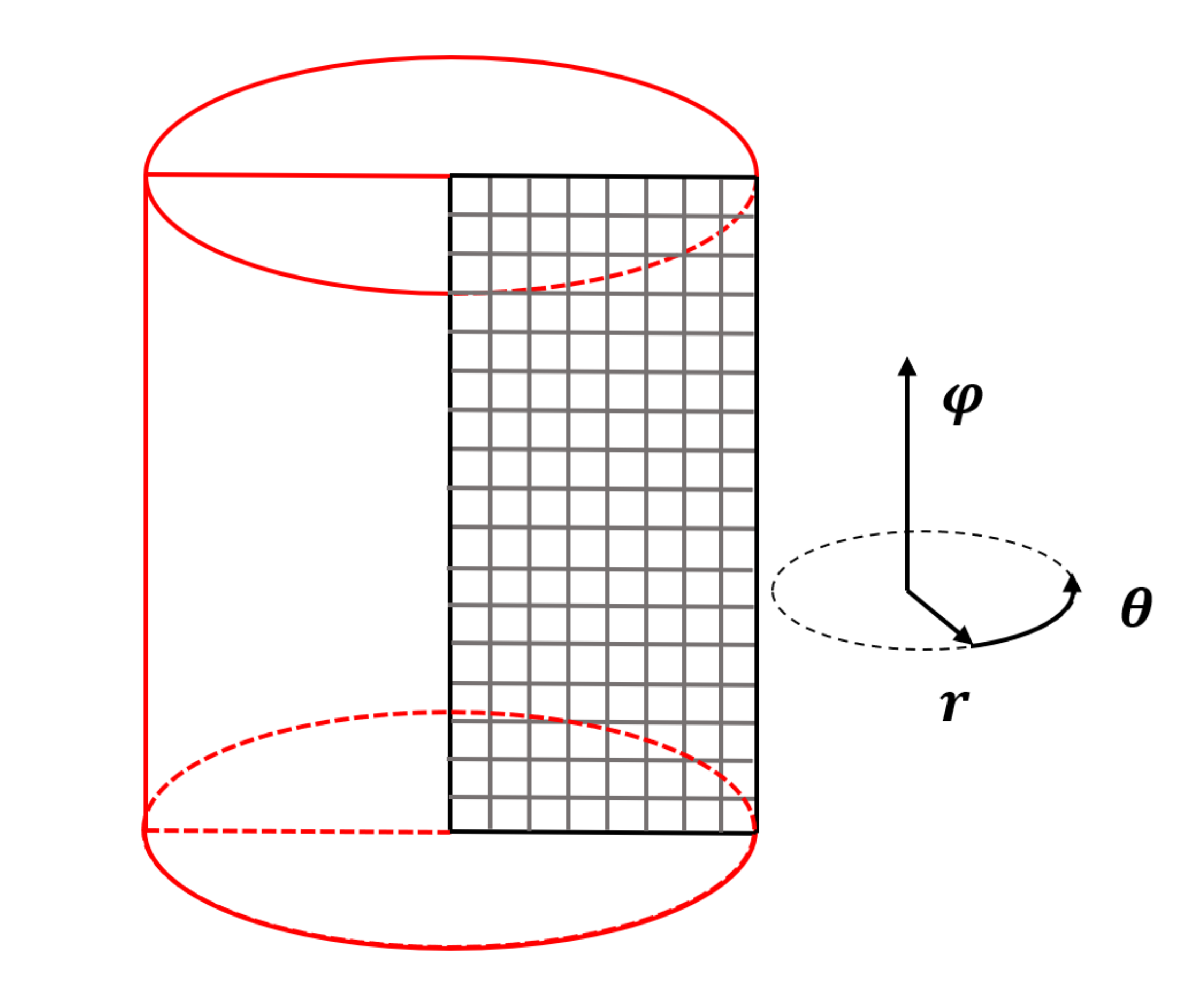}
\caption{The sketch of the 2D finite element grid and the 3D geometry used in the NIMROD simulation.  All fields are periodic in $\theta$ and $\varphi$ directions respectively.}  \label{fig:Grid}
\end{figure}

Since the  tearing modes with $m=-1, n\sim 2R/a$ are the dominant  instabilities in an RFP, the nonlinear simulation requires higher spatial resolution in the toroidal direction rather than poloidal direction. Therefore, a rectangular mesh composed of 32 by 64 (radial $\times$ toroidal) biquintic finite elements along with $6$ Fourier  components $0\leq |m| \leq 5$  in the poloidal direction  is used in the simulation in order to reduce the Fourier modes  required (Fig. \ref{fig:Grid}). The viscosity profile is given as $\nu(r)/\nu_0=[1+(\sqrt{19}-1)(r/a)^{20}]^2$, which is flat in the core region and   rises sharply only at the boundary to maintain numerical stability in the time advance.

\section{ Overview of a sawtooth event in simulation }
\begin{figure}[t]
\centering
\includegraphics[width=4in]{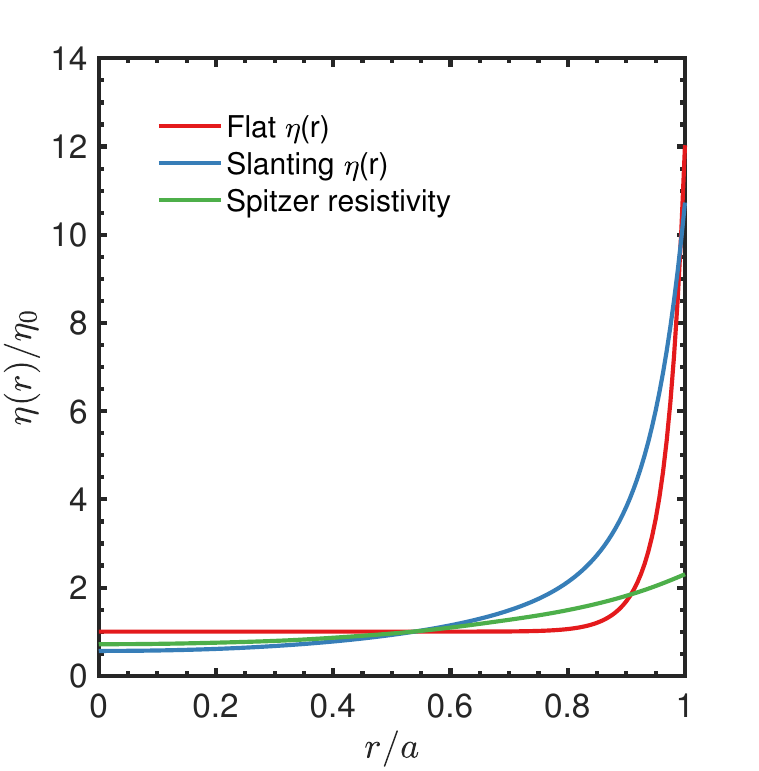}
\caption{Three different resistivity profiles as functions of the normalized minor radius used in the simulation. The Spitzer resistivity  is temperature dependent and varies over time, and the temporal averaged profile is shown.}  \label{fig:etaProfile}
\end{figure}

\begin{figure}[t]
\centering
\includegraphics[width=6in]{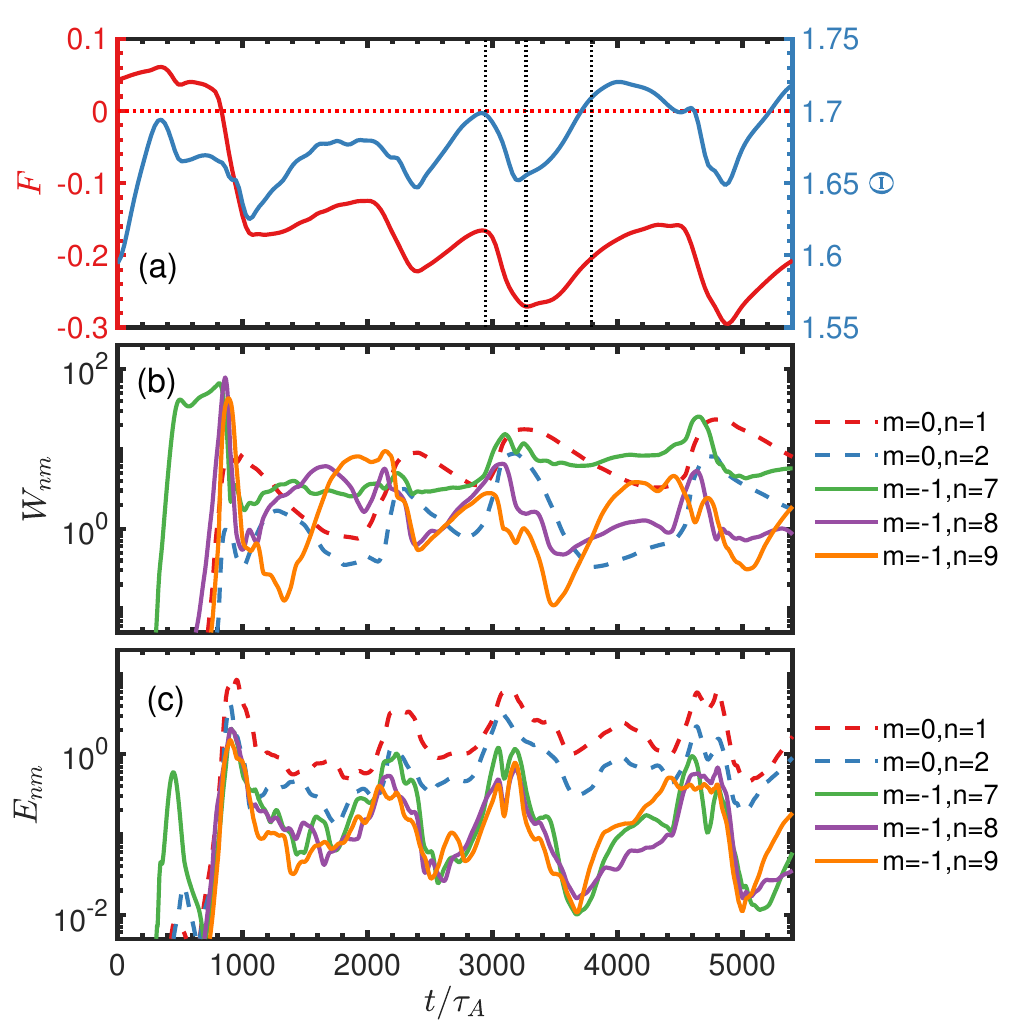}
\caption{(a) Field reversal parameter $F$ and pinch parameter $\Theta$, (b) magnetic energies and (c) kinetic energies of various $(m,n)$ mode components as functions of time  in the simulation.}  \label{fig:Overview}
\end{figure}

\begin{figure}[t]
\centering
\includegraphics[width=6in]{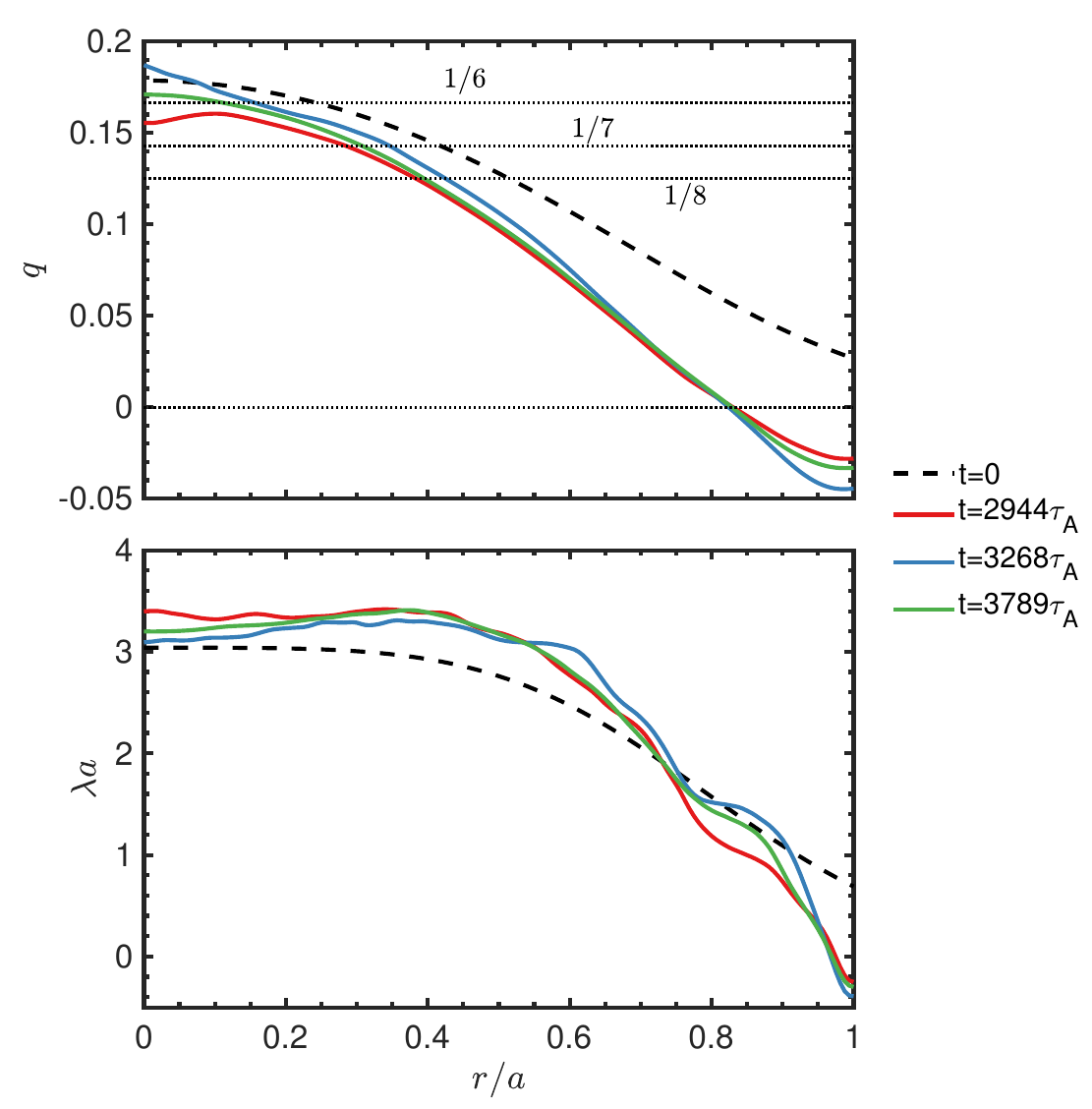}
\caption{(a) The safety factor profile  and (b) the dimensionless parallel  current  $\lambda$ profile at the initial state (i.e. t=0) and at t=2944$\tau_A$ before and t=3268$\tau_A$, 3789$\tau_A$ after a sawtooth event.}  \label{fig:Lambda}
\end{figure}

Based on the flat core resistivity profile in Fig. \ref{fig:etaProfile} with the parameters $S=M=2\times10^4$  in the core region, a typical quasi-periodic sawtooth oscillation  process obtained from the simulation is  outlined in Fig. \ref{fig:Overview}. Despite the  adopted $S$  level significantly lower than experimental values ($S>10^7$)  due to the limit of accessible computational resource,  the simulation reproduces most main features of the  sawtooth process often observed  in experiments. In particular, the most unstable tearing mode with $(m,n)=(-1,7)$ rapidly evolves to  saturation  and becomes  dominant  (Fig. \ref{fig:Overview}b-\ref{fig:Overview}c). This leads the plasma into a transient quasi single helicity (QSH) state \cite{Escande-2000} evolving from the initial paramagnetic pinch.
Subsequently, the emergence and growth of  secondary modes, such as $m=-1, n=8, 9$,  prompt  the plasma to exit QSH state and trigger  the nonlinear slinky mode, which excite the previously stable $m=0$ mode via three-wave coupling \cite{Craig-2017,Fitzpatrick-1999,Fitzpatrick-2002}. Once the nonlinear interaction surpasses a critical threshold, the plasma enters a rapid crash phase. During this phase, the amplitude of the primary mode decreases significantly, while the $m=0$ mode and other secondary modes grow further. Meanwhile, the magnetic reconnection event relaxes the peaked current profile, redistributing the plasma current from the interior to the exterior, particularly near the reversal surface, as illustrated in Fig. \ref{fig:Lambda}. The associated  magnetic flux pumping process is evident  in the $F$ and $\Theta$ parameters (Fig. \ref{fig:Overview}). As  the toroidal magnetic field drops to  a reversal state with $F \approx -0.2$,  $\Theta$ experiences a similar reduction.

After the initial sawtooth crash and reversal, the plasma repeats the sawtooth cycle quasi-periodically rather than remaining in the relatively low-energy relaxation state. The external drive  is not fully balanced by the plasma's self organization, and the current  profile  peaks again and the RFP plasma gradually recovers from the deep reversal state. The free energy provided by the current gradient excites instability, which leads to the subsequent sawtooth crash and rapid decrease of $F$ parameter.  Although the amplitudes of  these subsequent crashes are much smaller than the initial event due to the different start state, other sawtooth features do not exhibit significant difference. These characteristics are consistent with the observations of relaxation events in experiments and simulations \cite{Sauppe-2016,Momo-2020}.

\begin{figure}[t]
\centering
\includegraphics[width=6in]{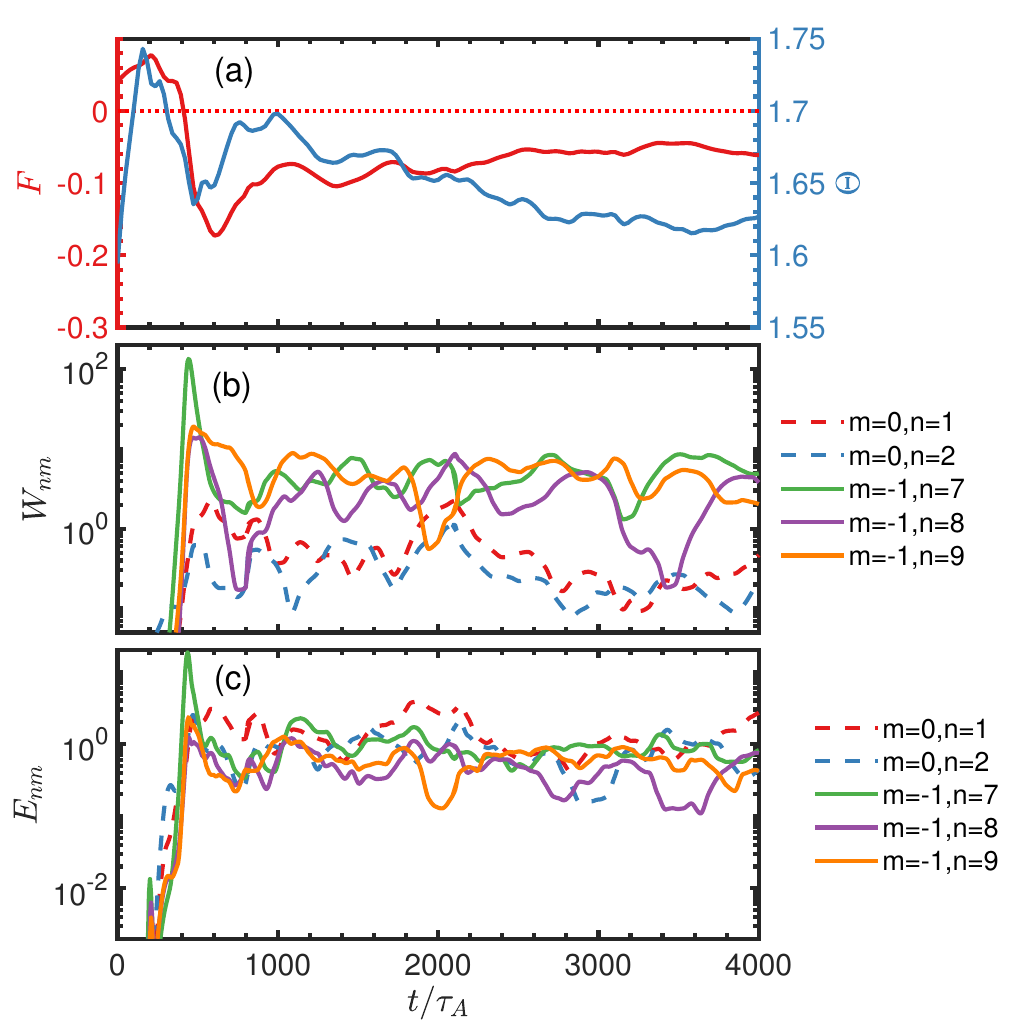}
\caption{(a) Field reversal parameter $F$ and pinch parameter $\Theta$, (b) magnetic energies and (c) kinetic energies of various $(m,n)$ mode components as functions of time  in the low $\Theta$ simulation in absence of sawtooth oscillation.}  \label{fig:Overview2}
\end{figure}

For a reduced  loop voltage, the sawtooth process in simulation exhibits  distinct dynamics (Fig. \ref{fig:Overview2}). After the initial sawtooth crash, the plasma maintains a shallow reversed field state with $F\approx -0.06$ without subsequent sawtooth oscillations, and the average $\Theta$ decreases from $1.68$ to  $1.63$ in comparison to the previous case.  In the absence of sawtooth process, the amplitude of the nonlinear $m=0$ mode diminishes significantly, and the toroidal reversal is  quasi-linearly maintained by the $m=-1$  tearing modes. The amplitudes of the tearing modes $m=-1,n=7,8,9$ are comparable  and exhibit temporal fluctuations. The disappearance of sawtooth can be attributed to the separation of reversal surface  and the resonance region as $\Theta$ decrease  \cite{Schnack-1985}. The Lundquist parameter $S$ is also known to influence the sawtooth process by enhancing nonlinear interactions, however the impact of resistivity inhomogeneity on the sawtooth process and dynamo field has been found significant as well in this work, which we elaborate in next section.

\section{Impact of non-uniform resistivity profile and dynamo effects}

In this section we present simulation results on the sawtooth process in  RFP with various resistivity profiles and analyze  the underlying physical mechanisms  from the perspective of dynamos.Three kinds of  resistivity $\eta$ profiles are used  for comparison (Fig. \ref{fig:etaProfile}). The first is flat in the core and steep near the edge, similar to the viscosity profile, which has been used in the simulation results reported in previous section. The second (``slanting'') type of $\eta$ profile is fitted from  experimental data on MST \cite{Reusch-2011}. The third kind  is calculated based on the  Spitzer resistivity model $\eta= \eta_0\times(T/T_0)^{-3/2}$, where $T_0$ and $\eta_0$  are the temperature and resistivity values at magnetic axis (i.e. r=0) respectively. To ensure that different resistivity profile yield the same $\Theta$  value under identical  loop voltage drive, there $\eta(r)$  profiles are not set same at $r=0$, with $\eta_{slanting}/\eta_{flat}|_{r=0}\approx1.7$. On the other hand, since the resistivity value at the boundary affects the dynamo balance  and the toroidal  magnetic field reversal \cite{Onofri-2010,Satherblom-1996},   the resistivity levels at the boundary are maintained  the same.

\subsection{Effects on the sawtooth process}

\begin{figure}[t]
\centering
\includegraphics[width=6in]{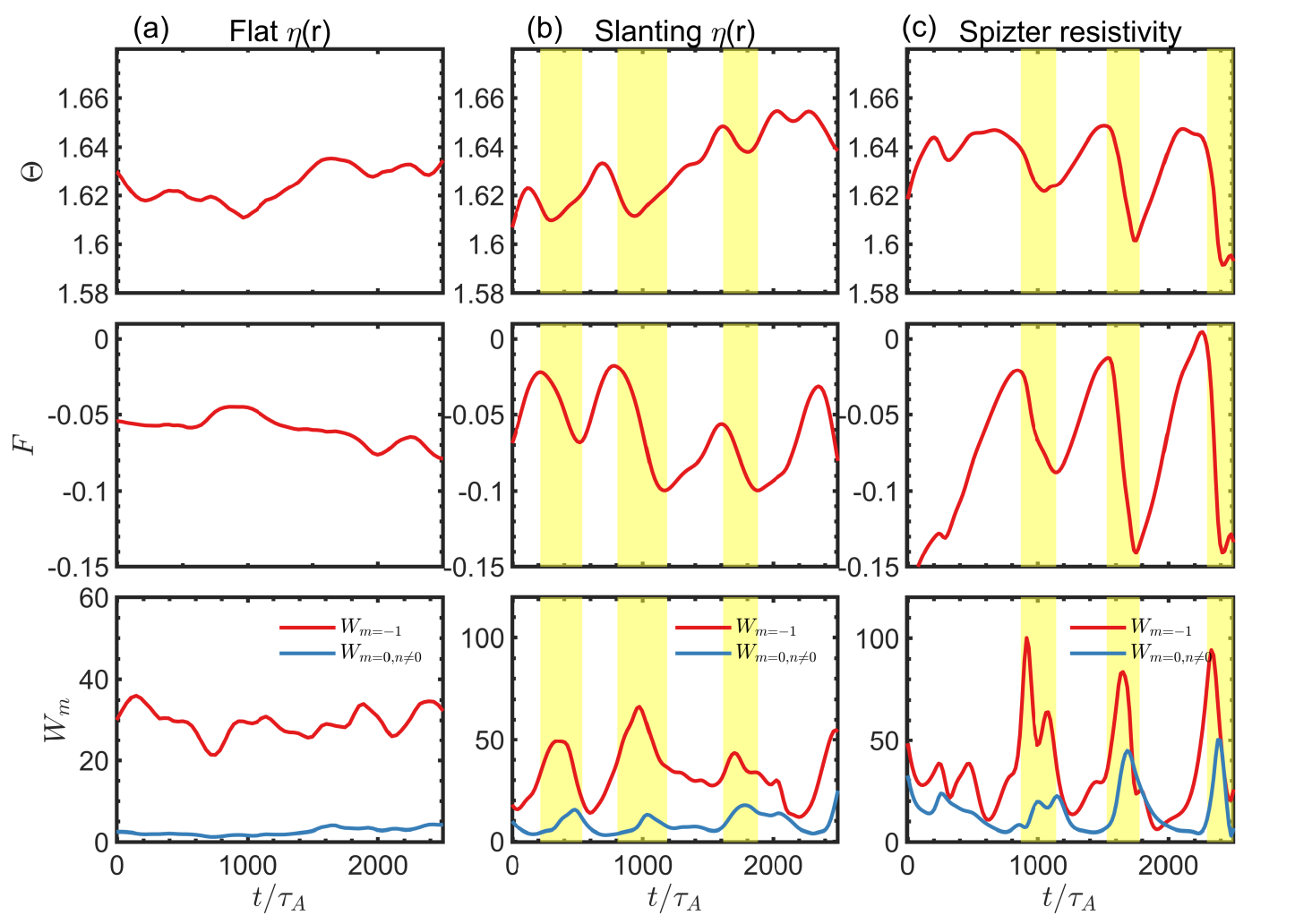}
\caption{The time evolution of pinch parameter $\Theta$, field reversal parameter $F$ and magnetic energy $W_{m}$ with various resistivity profiles: (a) flat resistivity profile,  (b) slanting resistivity profile and (c) Spitzer resistivity. In all three cases $S=2\times 10^4,M=2\times 10^4$. The sawtooth crash phases are highlighted in color yellow.}  \label{fig:VarProfileEvolution}
 \end{figure}

For three different resistivity profiles and the same toroidal voltage drive, the plasma currents are roughly the same with $\Theta\approx1.63$  (Fig. \ref{fig:VarProfileEvolution}). From the evolution of the $F$ parameter, it is evident that there is no sawtooth crash in the case with a flat resistivity profile, whereas clear sawtooth process appears in the cases with non-uniform resistivity profile and Spitzer resistivity.

\begin{figure}[t]
\centering
\includegraphics[width=6in]{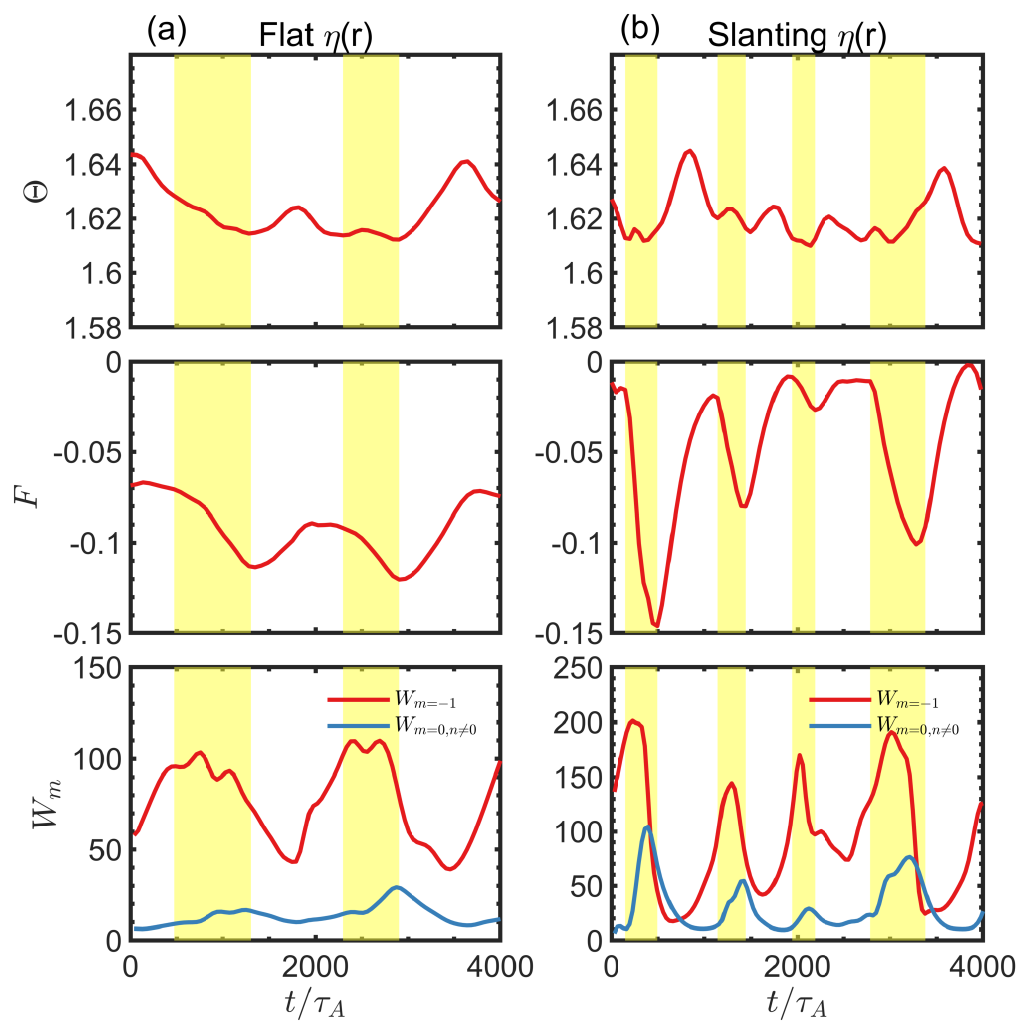}
\caption{The time evolution of pinch parameter $\Theta$, field reversal parameter $F$ and magnetic energy $W_{m}$ in simulation cases with various resistivity profiles: (a) flat resistivity profile,  (b) slanting resistivity profile. In both cases $S=8\times 10^4,M=2\times 10^4$. The sawtooth crash phases are highlighted in color yellow.}  \label{fig:VarSProfileEvolution}
 \end{figure}

The non-uniform resistivity profile also changes the existing sawtooth behavior (Fig. \ref{fig:VarSProfileEvolution}). When the resistivity is further reduced to $S=8\times 10^4$, periodic sawtooth events also occur in the case of a flat resistivity profile, with a small oscillation  $\delta F\approx 0.05$. When the resistivity profile is changed to a core-slanting profile, the period of sawtooth is reduced from approximately $1600\tau_A$ to $950\tau_A$, and the oscillation  amplitude increases to $\delta F\approx 0.12$. In this case, the average reverse field parameter of the plasma, $\langle F\rangle=-0.06$, is shallower than that in the case with a flat resistivity profile, $\langle F\rangle=-0.09$, which indicates that the former plasma case is further away from the fully relaxed state and has greater current free energy. Thus for the non-uniform resistivity profile, the nonlinear interaction during sawtooth bursts is stronger, as shown in the evolution of magnetic energies of modes $m=-1$ and $m=0$ in Fig. \ref{fig:VarSProfileEvolution}.

\begin{figure}[t]
\centering
\includegraphics[width=6in]{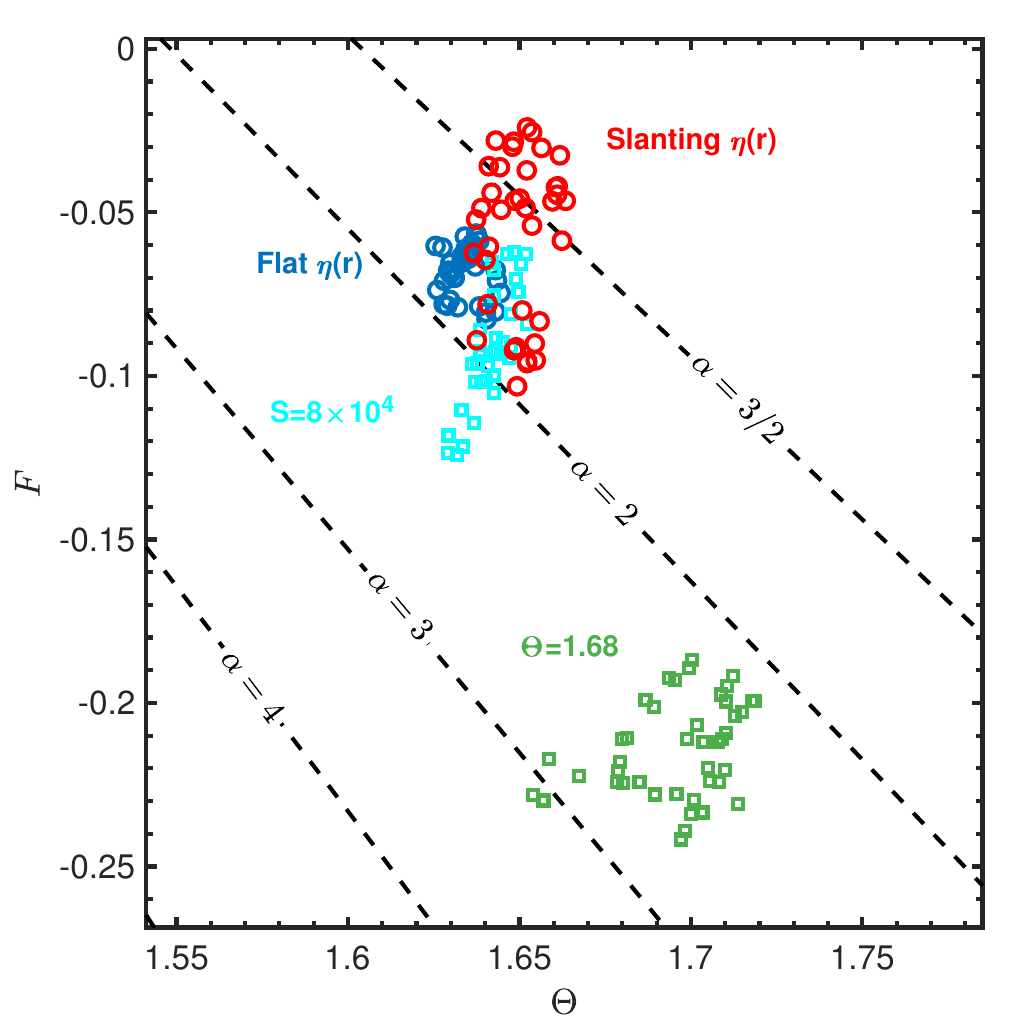}
\caption{The trajectories of sawtooth and sawtooth-free cases in the $F-\Theta$ phase diagram. The dashed lines represent the lines of constant $\alpha$  for  various $\lambda_0$ values.  The blue and red circles correspond to the flat and slanting resistivity profile cases  shown in Fig. \ref{fig:Overview2}, respectively. The green  and cyan squares represent the flat resistivity profile cases with a larger pinch parameter $\Theta=1.68$ and a larger Lundquist number $S=8\times 10^4$, respectively. }  \label{fig:FThetaDiagram}
 \end{figure}

The Modified Bessel Function Model (MBFM) is  used  for fitting RFP current profiles, where the  curl of magnetic field  \cite{Antoni-1986} can be written as :
\begin{equation}
\nabla \times \boldsymbol{B} = \lambda_0(1 - r^\alpha)\boldsymbol{B} + \left(\frac{\beta_0}{2B^2}\right)\boldsymbol{B} \times \nabla p
\end{equation}
where $\lambda_0=\mu_0 J/B$, and higher (lower) values of $\alpha$ correspond to a more uniform (peaked) current profile.
The $F-\Theta$ phase diagram in Fig. \ref{fig:FThetaDiagram} illustrates the transitions between  sawtooth oscillation and sawtooth-free states under various conditions.  The trajectory  of  the  sawtooth-free flat resistivity profile case, represented by the blue circles, is confined in a narrow region. In contrast,  the trajectory  of  the  slanting resistivity profile case, represented by the red circles, exhibits extensive oscillations in the phase diagram, primarily  corresponding to  variations in the $\alpha$ parameter. This corresponds to the process of alternating current profile peaking and relaxation during sawtooth oscillations. Furthermore, the non-uniform resistivity profile leads to  an overall reduction in the alpha parameter, which is also reflected in the shallower reversal parameter $F$ mentioned earlier. Furthermore, increasing the $\Theta$ parameter and the $S$ parameter can  induce a transition from a sawtooth-free state to a sawtooth state in RFP plasmas, as denoted by the blue and green markers in Fig. \ref{fig:FThetaDiagram}.

\subsection{Effects on dynamo electric fields}
To quantitatively describe the impact of non-uniform resistivity on the sawtooth process, we analyze the conditions required for the plasma to maintain a sawtooth free state\cite{Jardin-2020}. Starting from the time evolution equation of the magnetic vector potential $\boldsymbol{A}$ in the Coulomb gauge
\begin{equation}
\frac{\partial \boldsymbol{A}}{\partial t} = -\boldsymbol{E} - \nabla\Phi
\end{equation}
 Substituting the Ohm's law, and taking the toroidal direction projection, we obtain the following form of the magnetic flux evolution equation
\begin{equation}
-\frac{1}{2\pi R_0}\frac{\partial(\Psi_t+\Psi_{loop})}{\partial t} = \eta J_\phi - \boldsymbol{e}_\phi\cdot(\boldsymbol{v}\times\boldsymbol{B}) + \boldsymbol{e}_\phi\cdot\nabla\Phi
\end{equation}
where $\Psi_t$ is  the toroidal flux in plasma other than the toroidal flux $\Psi_{loop}$  generated by loop voltage, $\Phi$ represents the electrostatic potential, and the rest of symbols are conventional. After taking the surface  average $\langle f\rangle=\int{}d\theta{}\int{}d\phi{}f/(2\pi)^2$ and keeping only the zeroth-order components, the electrostatic potential term vanishes, and we obtain the evolution equation for the toroidal magnetic flux as
\begin{equation}
-\frac{1}{2\pi R_0}\frac{\partial \Psi_t}{\partial t} = \langle \eta J_\phi \rangle - \langle \boldsymbol{e}_\phi\cdot(\boldsymbol{v}\times\boldsymbol{B}) \rangle - E_{loop}
\end{equation}
Similarly, taking the poloidal direction projection yields the evolution equation for the poloidal magnetic flux
\begin{equation}
-\frac{1}{2\pi r}\frac{\partial \Psi_p}{\partial t} = \langle \eta J_\theta \rangle - \langle \boldsymbol{e}_\theta\cdot(\boldsymbol{v}\times\boldsymbol{B}) \rangle
\end{equation}

It is evident that for a plasma to achieve and sustain a steady state  for the toroidal and poloidal magnetic fluxes, the sum of the dynamo electric field term,  the resistive term, and  the loop voltage term on the right side must also equal to zero (there is no external electric field in the poloidal direction). Among the three terms on the right side of the equation, the loop voltage term is determined by the external driving electric field, the resistive term $\langle\eta\boldsymbol{J}\rangle$ is determined by the resistivity profile and the equilibrium current, and the fluctuation induced dynamo electric field reflects the self-organization behavior of the plasma, with contributions from multiple modes.

\begin{figure}[t]
\centering
\includegraphics[width=6in]{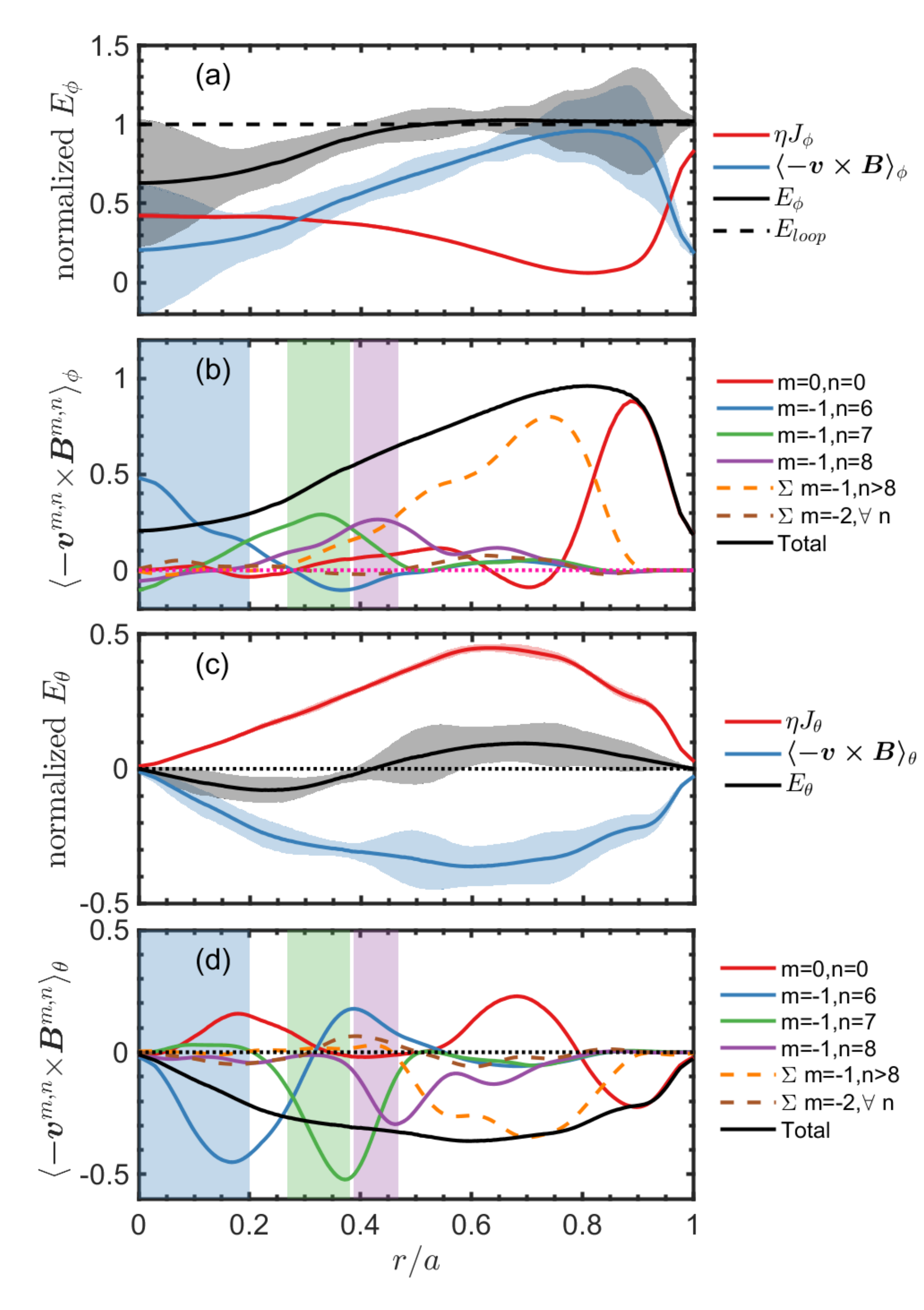}
\vskip -0.3in
\caption{(a) The toroidal components  and (c) poloidal components in the generalized Ohm's equation during the rising phase of the sawtooth cycle, where lines and shaded regions
represent the average value and the range of  electric field at various times, respectively. Modal decomposition of  (b) toroidal and (d) poloidal  dynamo electric fields, where the shaded regions with various colors illustrate  the radial positions of the resonance surfaces for the corresponding mode.}  \label{fig:RiseDynamo}
 \end{figure}

\begin{figure}[t]
\centering
\includegraphics[width=6in]{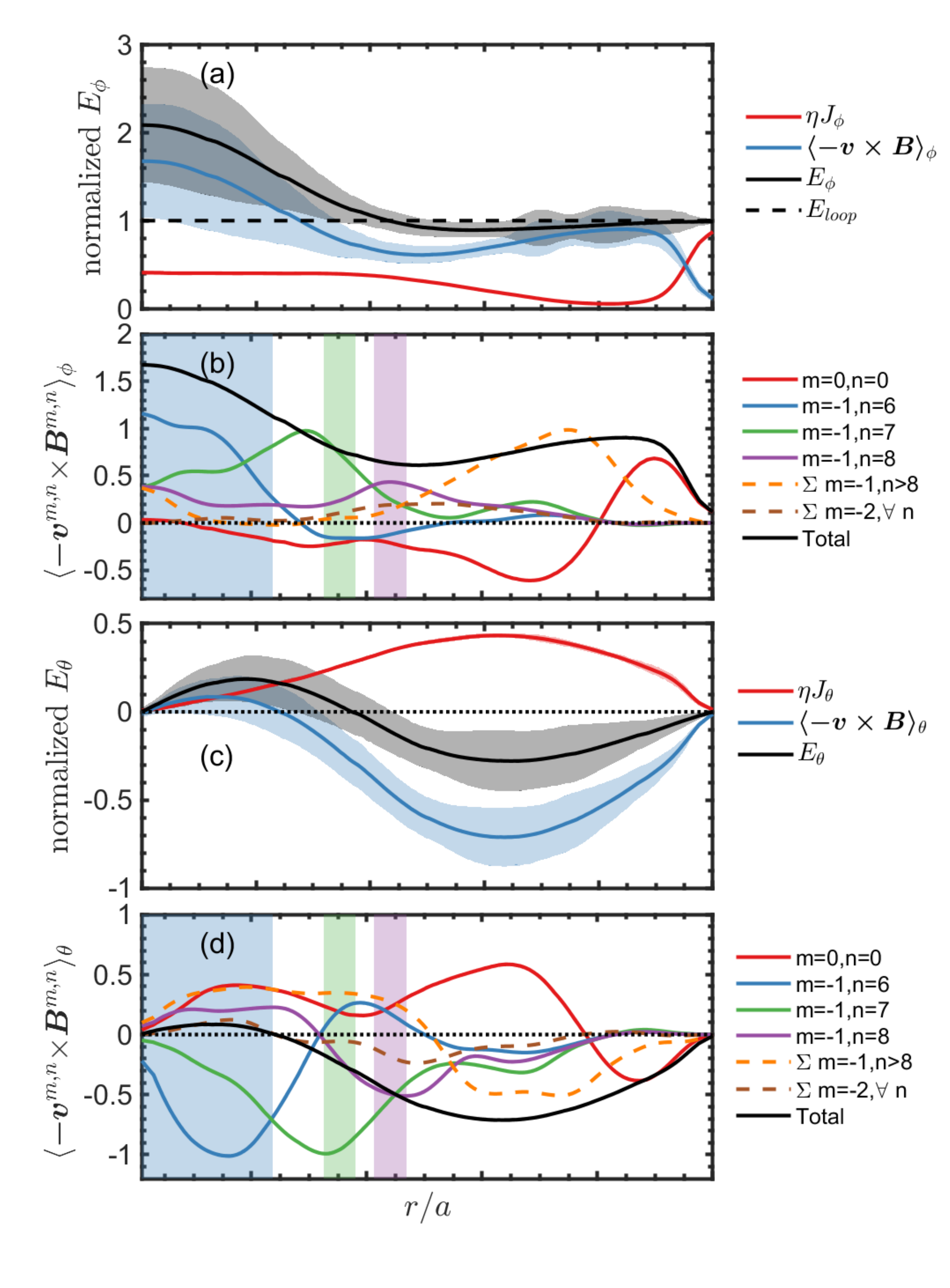}
\caption{(a) The toroidal components  and (c) poloidal components in the generalized Ohm's equation during the crash phase of the sawtooth cycle, where lines and shaded regions
represent the average value and the range of  electric field at various times, respectively. Modal decomposition of  (b) toroidal and (d) poloidal  dynamo electric fields, where the shaded regions with various colors illustrate  the radial positions of the resonance surfaces for the corresponding mode.}  \label{fig:CrashDynamo}
 \end{figure}

The toroidal and poloidal components in the  Ohm's law during the rising and crash phases of the sawtooth cycle have been time-averaged and normalized with the loop voltage electric field $E_{loop}$ (Figs. \ref{fig:RiseDynamo} and \ref{fig:CrashDynamo}). As discussed above, non-zero net electric field  induces  time variations in the magnetic field. During the rising phase, the net toroidal electric field $E_\phi-E_{loop}$ is negative  inside and positive outside of the radius $r/a=0.53$, driving counter-currents  and co-currents in the corresponding regions respectively, and transporting the poloidal magnetic flux from the outside region to the core.  The poloidal electric field behaves similarly, which is  negative inside and positive outside of the radius $r/a=0.42$, transporting the toroidal magnetic flux from the outside region to the core. These dynamics  leads to the current peaking process during the sawtooth rising phase, with the toroidal dynamo effect being more dominant. Conversely, during the crash phase, the scenario reverses. With little change in the external driving electric field and dissipative electric field, the major change in the dynamo electric field enables the total electric fields in both toroidal and poloidal directions to switch signs from negative inside and positive outside to positive inside and negative outside of the transition radius. The amplitude of the net electric field is much higher than the rising phase, corresponding to the rapid pumping of magnetic flux from the core to the outside during the crash phase of magnetic reconnection.

During  the  rising phase,  the dynamo electric fields generated  by  different  $m=-1$ tearing modes are mainly concentrated near their respective resonance surfaces  at $r/a\approx0.12, 0.32, 0.42$ for $n=6, 7, 8$ (Fig. \ref{fig:RiseDynamo}b).  As  the $n$ mode number increases  from low to high, the peak locations of the dynamo electric field profile for the corresponding modes  shift outward from the interior to the edge. The $m=0$ mode's contribution is similarly localized near the reversal surface. The $m=-2$ mode arises from the nonlinear interaction among $m=-1$ modes, exhibiting a dynamo electric field with a comparatively reduced amplitude, peaking between the reversal and core resonant surfaces. The  dynamo electric field during the crash phase share similar spatial distribution characteristics except for  the significantly larger amplitude.

\begin{figure}[t]
\centering
\includegraphics[width=6in]{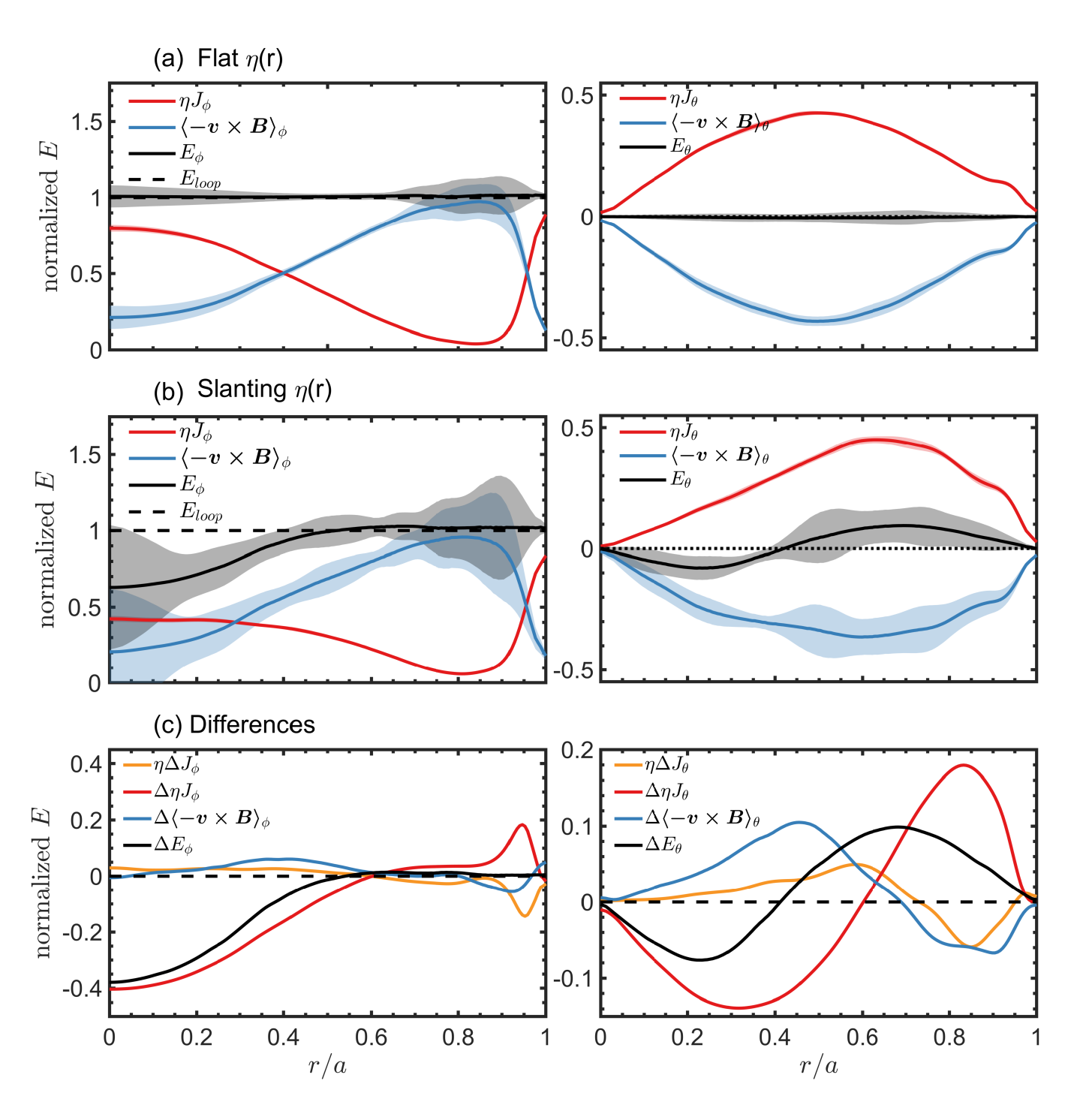}
\caption{(a) The toroidal   and  poloidal components in the generalized Ohm's law in the sawtooth-free case with a flat resistivity profile; (b)  the same components  during the sawtooth rising phase  with a slanting resistivity profile;  (c) the differences of the components between the above two rows.}  \label{fig:VarProfileDynamo}
 \end{figure}

In the case of a flat resistivity profile, the plasma state does not exhibit periodic oscillations but remains relatively steady. As illustrated  in Fig. \ref{fig:VarProfileDynamo}a, there are nearly zero net toroidal and poloidal electric fields. To better elucidate  the influence of non-uniform resistivity profiles on the plasma during the resistive diffusion phase, the electric fields between different resistivity profile cases are subtracted, and $\Delta\boldsymbol{E}$ is divided into the following three terms

\begin{equation}
\Delta \boldsymbol{E} = \Delta\eta\boldsymbol{J}_{n=0} + \eta\Delta\boldsymbol{J}_{n=0} + \Delta\langle-\boldsymbol{v}\times\boldsymbol{B}\rangle
\end{equation}
where  the alteration in  resistivity profile is denoted as $\Delta\eta$, the change in  current profile is represented as $\Delta\boldsymbol{J}$, and the resulting variation of plasma dynamo is written as $\Delta\langle-\boldsymbol{v}\times\boldsymbol{B}\rangle$  (Fig. \ref{fig:VarProfileDynamo}c). In both toroidal and poloidal directions, $\Delta\eta\boldsymbol{J}$ is the main contribution to the difference in electric fields,  which in turn determines the electric field profile's characteristic structure, i.e., negative near the core and positive toward the outside. The peaking of plasma current profile is associated with the $\eta\Delta\boldsymbol{J}$ electric field. In comparison to the flat resistivity profile case (Fig. \ref{fig:VarProfileDynamo}a), in the presence of a non-uniform resistivity profile  (Fig. \ref{fig:VarProfileDynamo}b), the toroidal dynamo electric field $\langle-\boldsymbol{v}\times\boldsymbol{B}\rangle_\phi$  and the poloidal dynamo electric field $\langle-\boldsymbol{v}\times\boldsymbol{B}\rangle_\theta$  are enhanced in most regions but weakened near the boundary, thereby partially offsetting the impact of the resistivity profile change.

\subsection{Source of the dynamo electric field}

\begin{figure}[t]
\centering
\includegraphics[width=5in]{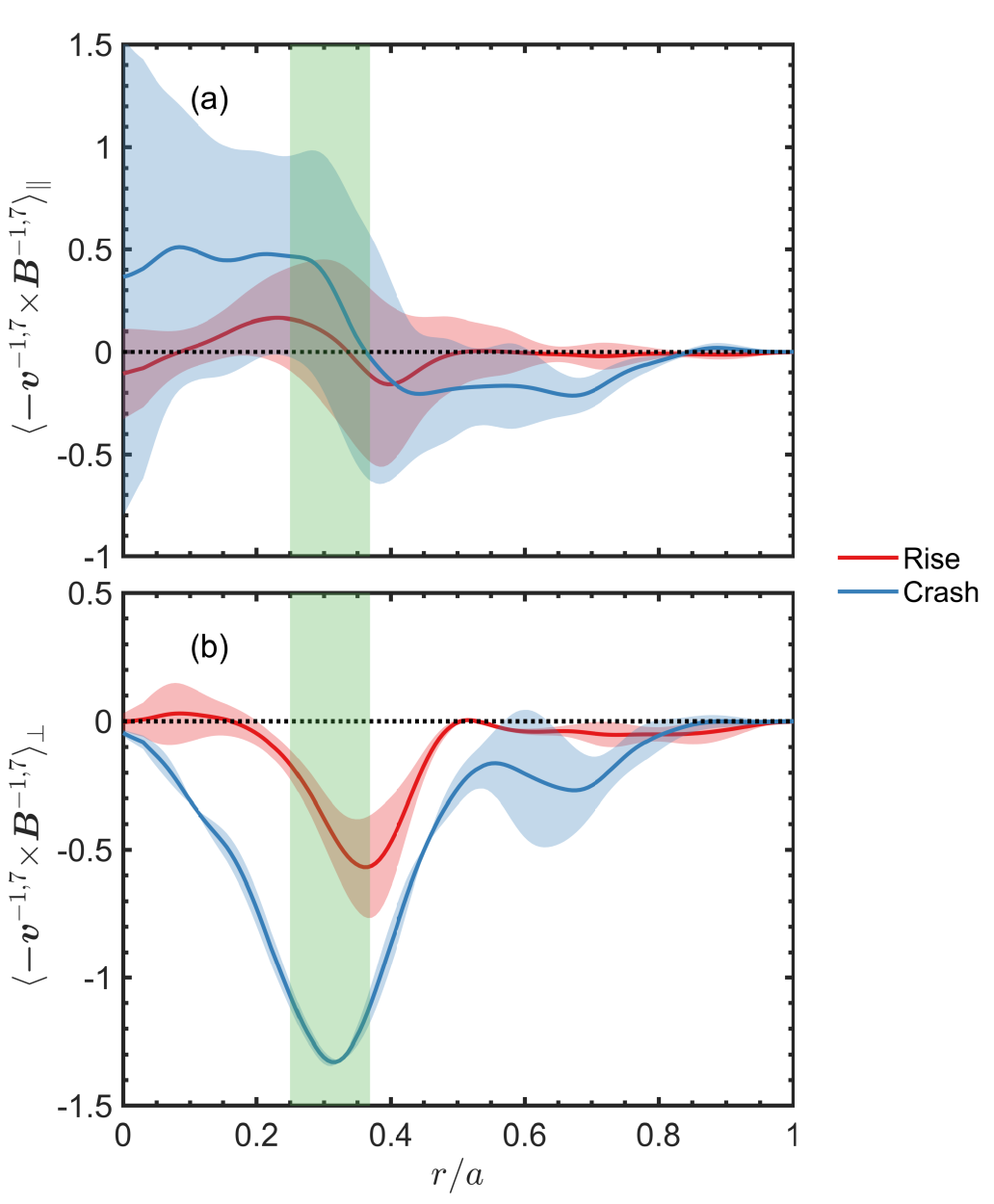}
\caption{The radial profile of dynamo electric field generated by the $(-1,7)$ mode in  (a) parallel and (b) perpendicular  directions. The red and blue lines represent the electric fields during the sawtooth rising and crash stages, respectively. Lines and shaded regions represent the average value and range of the electric field at various times, respectively, where the radial position of the $(-1,7)$ mode's resonance surface is illustrated using the green shaded region.}  \label{fig:Dynamo17}
 \end{figure}

Fig. \ref{fig:RiseDynamo} and Fig. \ref{fig:CrashDynamo} demonstrate the principal contributions of the $m=-1$ and $m=0$ modes to the dynamo electric fields. The dynamo electric field of the $m=0$ mode predominantly arises from the radial velocity perturbation ${v_r}$, which generates a dynamo field   perpendicular to the equilibrium magnetic field. The flow field associated with the $m=-1$ component of dynamo electric field is notably more complicated. Thus, the dynamo electric field generated by the specific  mode $(-1,7)$ throughout  the sawtooth rising and crash phases has been time-averaged and decomposed into components perpendicular and parallel to the magnetic field. These results are presented in Fig. \ref{fig:Dynamo17}. It is evident that the parallel component of the $(-1,7)$ mode's dynamo electric field is negligible  during the sawtooth rising but significantly contributes to the crash phase, exhibiting as positive inside and negative outside of the resonance surface.  In contrast, the perpendicular component consistently provides  a negative contribution  across  both phases,  reaching its peak  near the resonance surface. The subsequent analysis focuses on the vertical and parallel electric fields during the crash phase.

For the purpose of visualizing the spatial distribution of electromagnetic and flow fields, physical quantities with single perturbation are projected onto a helical surface.
For convenience, $\tilde{\boldsymbol{v}}$ and $\tilde{\boldsymbol{b}}$ are used in the following to refer to the $(-1,7)$ Fourier components of  $\boldsymbol{v}$ and $\boldsymbol{B}$ respectively. Let ${\boldsymbol{e}_r}$ denote  the normal direction to the helical surface, $\boldsymbol{e}_\zeta=(-\boldsymbol{e}_\phi m/r+\boldsymbol{e}_\theta n/R_0)/l$, and  $\boldsymbol{e}_\xi=\boldsymbol{e}_\zeta\times\boldsymbol{e}_r=(-\boldsymbol{e}_\phi n/R_0-\boldsymbol{e}_\theta m/r)/l$, with $l=\sqrt{(m/r)^2+(n/R_0)^2}$ and $\xi=-m\theta-n\phi$ (note that m<0 here). Proximity to the $(m,n)$ resonance surface enables the unit vectors $\boldsymbol{e}_\zeta$ and $\boldsymbol{e}_\xi$ to serve as approximations for the  parallel and the cross directions to the equilibrium magnetic field, respectively.

\begin{figure}[t]
\centering
\includegraphics[width=4in]{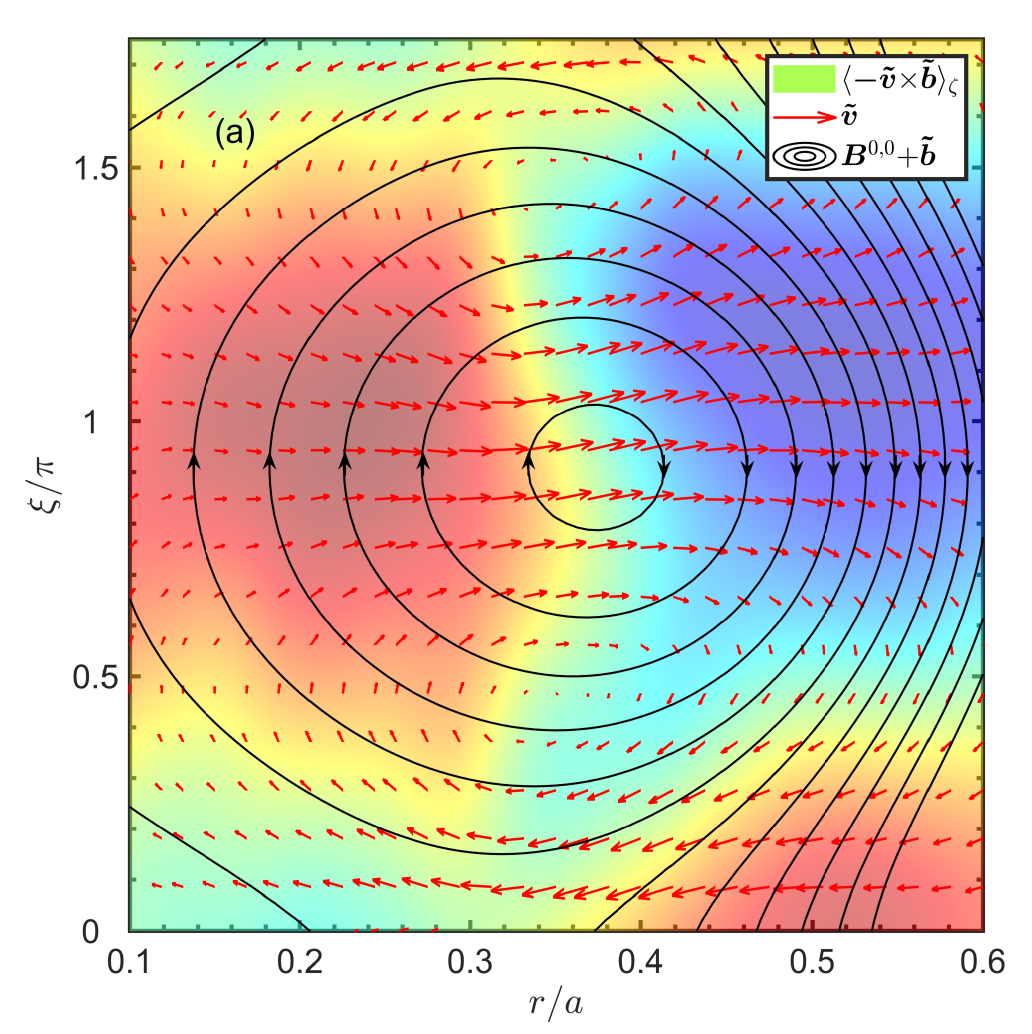}
\includegraphics[width=4in]{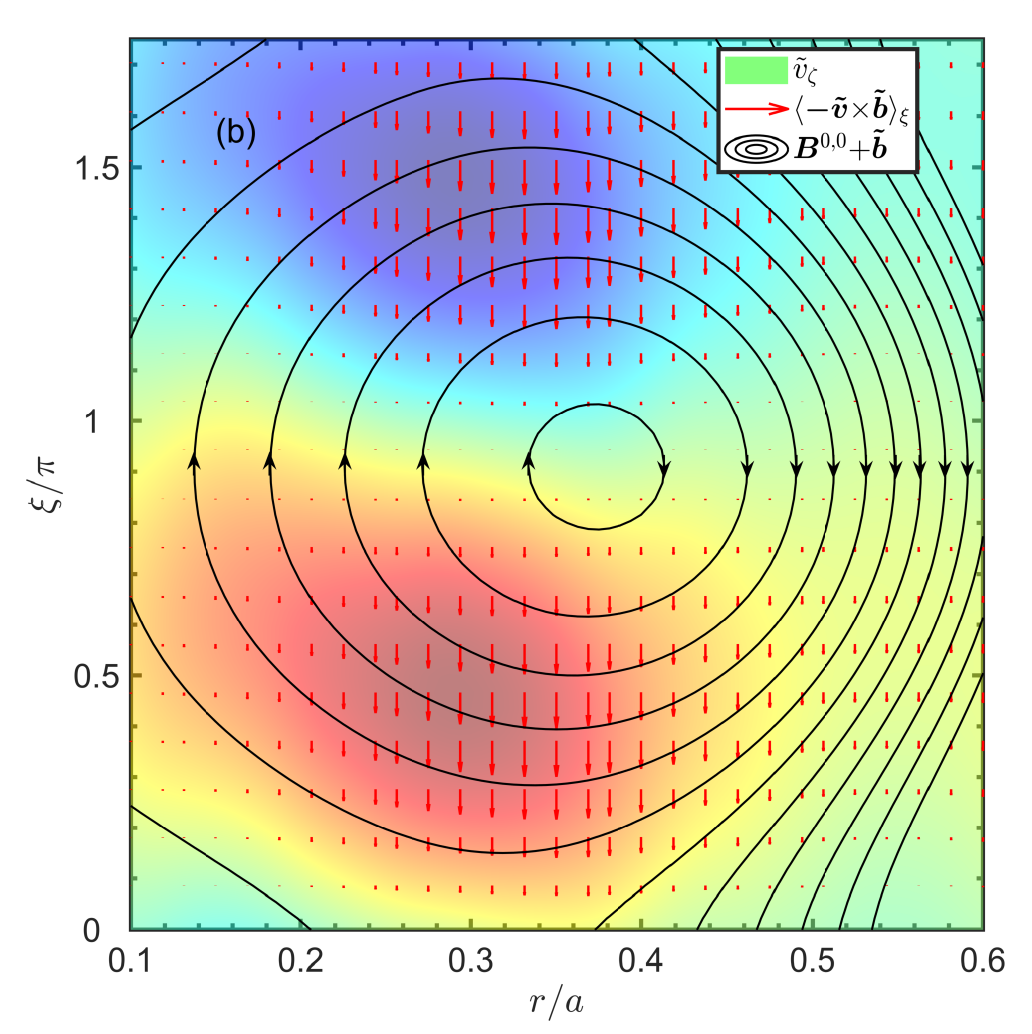}
\caption{(a) Helical projection of the $(-1,7)$ perturbed flow $\boldsymbol{\tilde{v}}_\zeta $ and the corresponding dynamo field in the parallel direction $\langle -\boldsymbol{\tilde{v}}\times \boldsymbol{\tilde{b}}\rangle _\zeta$, and (b) helical projection of the $(-1,7)$ perturbed parallel flow $\tilde{v}_\zeta$ and the corresponding of dynamo field in the cross direction $ \langle \boldsymbol{\tilde{v}}\times \boldsymbol{\tilde{b}}\rangle _\xi =\langle \tilde{b}_r\times \tilde{v}_\zeta-\tilde{b}_\zeta\times\tilde{v}_r\rangle\approx \langle \tilde{b}_r\times \tilde{v}_\zeta\rangle$ }.
  \label{fig:HelicalFlow}
 \end{figure}

The spatial structure of the dynamo electric field at the reconnection site during the sawtooth crash phase can be illustrated using the helical projections of its parallel and cross components (Fig. \ref{fig:HelicalFlow}). The parallel component of the dynamo electric field, resulting from the sum of $\tilde{v}_\xi\times \tilde{b}_r$ and $-\tilde{v}_r\times \tilde{b}_\xi$, is positive inside and negative outside  the resonance radius. The radial flow $\tilde{v}_r$ emanates outward at the O-point and inward at the X-point of the magnetic island; meanwhile, the $\tilde{v}_\xi$ component flows from the X-point to the O-point inside  the resonance radius but switches its direction outside. \mbox{Fig. \ref{fig:HelicalFlow} b} illustrates the negative dynamo field in the cross direction during the crash phase. The flow component $\tilde{v}_\zeta$  is in phase with the radial magnetic field $\tilde{b}_r$, and their product  $-\tilde{v}_\zeta\times\tilde{b}_r $ constitutes a primary contribution to the cross component of dynamo field.

\section{Conclusion and discussion}

This paper thoroughly  investigate  the plasma sawtooth dynamics  in three-dimensional simulations under Ohmic driving with different resistivity profiles. The findings indicate that plasma current profiles are more susceptible to peaking under non-uniform resistivity profiles, thereby  providing conditions for the periodic  occurrence of sawtooth relaxation events. Moreover, the non-uniform resistivity profile also modulates  the dynamo electric field required to sustain the reversed field in the plasma, making the reversed field shallower.

We further analyze the impact of non-uniform resistivity profiles on the dynamo electric field. By comparing simulation results across different resistivity profiles, we find that during the sawtooth rising phase, the variation in the resistivity profile is the primary contributor to the differences in the electric field, driving the peaking of the plasma current, and the acceleration  of the inward diffusion of magnetic flux. The additional resistive diffusion effect brought by the non-uniform resistivity profile ultimately increases the reconnection amplitude during the sawtooth crash phase, significantly enhancing the dynamo electric field and the magnetic flux pumping effect. Furthermore, we explore and find out how the flow field of the $m=-1$ mode during the crash phase generates the dynamo electric field in both perpendicular and parallel directions.

In the  temperature dependent resistivity case,  the plasma $\beta$ at  edge rises to approximately 5\% due to Ohmic heating, which has also been adopted for all other cases in the study. However,  for the higher $\beta$  regime beyond the current study,  two-fluid effects such as FLR (Finite Larmor Radius) and Hall dynamo may no longer be ignored. Previous two-fluid simulations of RFP \cite{King-2011,Sauppe-2017} indicate that the Hall dynamo may impede the MHD dynamo during relaxation events, even though the overall dynamo effect may be similar to single-fluid model. In addition to two-fluid effects, the $\beta$ parameter  is also known to have direct influences on the self-organization and dynamo response process \cite{Zhang-2020,Luo-2017}. Future work is planned to study the dynamo drive for RFP sawtooth relaxation process in the higher $\beta$ regimes using the two-fluid MHD model.
\section*{Acknowledgment}

The authors express sincere thanks to all members of the KTX group as well as the NIMROD team. This work was supported by the National Natural Science Foundation of China (Grant No. 11775220), the National Magnetic Confinement Fusion Energy Development Program of China (Grant Nos. 2017YFE0301702 and 2019YFE03050004), and U.S. Department of Energy (Grant No. DE-FG02-86ER53218). The computing work in this paper was supported by  the Supercomputing Center of USTC and the Public Service Platform of High Performance Computing by Network and Computing Center of HUST.

\newpage
\section*{References}

\bibliographystyle{unsrt}
\bibliography{References} 
\newpage

\newpage

\newpage

\end{document}